\documentclass[aip,reprint,nofootinbib]{revtex4-1}
\usepackage{bm}

\begin{document}
\title{The Final Theory of Physics - a Tautology?}
\date{\today}
\author{C. Baumgarten}
\affiliation{5244 Birrhard, Switzerland}
\email{christian-baumgarten@gmx.net}

\def\begeq{\begin{equation}}
\def\endeq{\end{equation}}
\def\begary{\begeq\begin{array}}
\def\endary{\end{array}\endeq}
\def\bmtx{\left(\begin{array}}
\def\emtx{\end{array}\right)}
\def\eps{\varepsilon}
\def\d{\partial}
\def\y{\gamma}
\def\w{\omega}
\def\W{\Omega}
\def\s{\sigma}
\def\ket#1{\left|\,#1\,\right>}
\def\bra#1{\left<\,#1\,\right|}
\def\bracket#1#2{\left<\,#1\,\vert\,#2\,\right>}
\def\erw#1{\left<\,#1\,\right>}

\def\Exp#1{\exp\left(#1\right)}
\def\Log#1{\ln\left(#1\right)}
\def\Sinh#1{\sinh\left(#1\right)}
\def\Sin#1{\sin\left(#1\right)}
\def\Tanh#1{\tanh\left(#1\right)}
\def\Tan#1{\tan\left(#1\right)}
\def\Cos#1{\cos\left(#1\right)}
\def\Cosh#1{\cosh\left(#1\right)}
\def\FiT{\textsc{FiT }}
\def\GUT{\textsc{GUT }}

\begin{abstract}
We acuminate the idea of a final theory of physics in order to analyze its 
logical implications and consequences. It is argued that the rationale of a
final theory is the principle of sufficient reason. This implies that a final 
theory of physics, presumed such a theory is possible, does not allow to 
incorporate substantial (non-trivial) propositions unless they are logically
or mathematically deduced. Differences between physics and mathematics are
discussed with emphasis on the role of physical constants. It is shown that 
it is logically impossible to introduce constants on the fundamental level 
of a final theory. The most fundamental constants emerging within a final
theory are constants of motion.

It is argued that the only possibility to formulate a final theory 
is necessarily a tautology: A final theory of physics can only be derived
from those presumptions about reality that are inherent in the idea and
practice of physics itself. It is argued that a final theory is based on
the notion of objectivity, but it is logically impossible that an ideal
final theory supports realism. 
\end{abstract}
\maketitle

\section{Introduction}
\label{sec_intro}

In his book {\it Dreams of a Final Theory} Nobel laureate Steven Weinberg
wrote~\cite{Weinberg}: ``It is difficult to imagine that we could ever be 
in possessions of final physical principles that have no explanation in 
terms of deeper principles.'' But Weinberg (of course) did exactly this:
he described the {\it idea of a final theory} in his book. 
Not all physicists, not even all Nobel laureates, share Weinbergs expectation 
that such a theory might be possible~\cite{ToE2005}. Nevertheless we shall 
pick up and discuss various aspects of Weinbergs dream. Specifically we shall
use the idea of the final theory to discuss the relation between physics
and mathematics. Whether we believe or not that a final is possible 
{\it in principle}, both cases have their specific implications~\footnote{
It might be the historical fate of physics to fail providing a final theory, 
it might end up with a {\it last} instead of a {\it final} theory - but we 
will not consider this possibility.}.
Weinberg's dream concerns a theory that is final {\it because no deeper theory 
is possible}. The concept of a {\it final} theory (\FiT) understood in this
way differs considerably from that of a {\it grand unified theory} (\GUT) or a 
{\it theory of everything}. A \GUT aims to provide a {\it unified} description 
of all fundamental forces of nature. However revolutionary this would be, it 
remains within the scope of already existing theories; the concept of a \GUT 
does not necessarily imply {\it finality}, but {\it completeness} with respect
to known experimental facts. 

The conjectured finality results, as Weinberg explains, from ``the beauty of 
simplicity and inevitability, the beauty of perfect structure, the beauty of 
everything fitting together, of nothing being changeable, of logical rigidity.''
The keyword here is {\it inevitability}: Though the idea of a \GUT 
implies {\it completeness} with respect to the known fundamental particles and
forces, it neither implies simplicity, nor a deep or even final level of understanding, 
nor a high degree of explanatory power. A \GUT could be ugly and incomprehensible.
The form of a \GUT - if we extrapolate the prominent tendencies in theoretical 
physics - will likely consist of a set of mathematical structures that allow 
to summarize the structure of the standard model of particle physics. 
But the very idea of a \GUT does not include explanations why nature 
should prefer a specific mathematical structure and not some other.
Or, in other words, a \GUT would - in contrast to the ideal \FiT - not exclude 
the possibility that different universes with different laws of nature might
exist at least in principle.

However, we shall argue that the conjectured {\it finality} of a \FiT
follows a different logic than the assumed completeness of the \GUT.
Furthermore we shall argue that and why Weinberg's specifications can, 
if at all, only be met by a theory that is based on {\it trivial} premises. 
The finality and inevitability that Weinberg expects from a final theory 
implies that it is similar to pure mathematics. This has a number of 
consequences~\footnote{One implication of the inevitability 
seems to be that the final theory is incompatible with the 
``Mathematical Universe Hypothesis'' in Tegmark's sense of enumerable 
possible universes~\cite{Tegmark}: If many worlds are possible, then 
none of them can be regarded as inevitable.}.

Weinberg wrote ``But (I repeat and not for the last time) I am concerned 
here not so much with what scientists {\it do} [...] as I am with the logical 
order built into nature itself''. We we shall argue that a final theory 
(if possible at all) will be based exactly on this: on what physicist {\it
  do}, what they {\it have to do}, in order to formulate a quantitative 
and objective description of nature.

In this essay we further elaborate the premises that we used in two preceeding
publications~\cite{qed_paper,osc_paper} in which the author sketched a logical
line of reasoning that allows to derive relativity, electrodynamics and
the Dirac equation from (almost) trivial principles. Here we shall argue 
that these principles are a promising starting point of a final theory - 
not because we are able to deduce a \GUT from it, but because they scetch 
a possibility to achieve the required logic of a \FiT.

We shall argue that specifically the inevitability and finality of a 
physical theory can only be derived from a systematic analysis of the 
{\it form} or {\it method} of physics itself: Since the theory has to 
demonstrate that any physical world inevitably has the form of our universe, 
then it can not be purely empirical: it has to demonstrate logical necessity. 
And since we do (like Weinberg) not believe that it can be derived 
``from pure thought'', then it can only be derived as an tautology, 
i.e. from a description of physics itself.

In Sec.~\ref{sec_fit} we will try to precisely specify what we consider to
be a final theory of physics. We do not know if our definition is in
full agreement with Weinbergs idea of a \FiT and his attitude towards 
such a theory. We will acuminate the idea to an extend that allows to draw
conclusions. Insofar it is a thought's experiment.

In Sec.~\ref{sec_math} we discuss aspects of the relation and the differences 
between mathematics and physics and the consequences for a \FiT. Here we put
special emphasis on the difference in the methods of concept formation 
between mathematics and physics and what can be concluded for a final theory.

In Sec.~\ref{sec_ref} we investigate the role played by the so-called 
``physical constants'', both, dimensional and dimensionless. We shall discuss 
where and how such constants may be introduced or derived within a final theory.
The specific meaning of natural units is considered in Sec.~\ref{sec_natural_units},
their role within physical laws in Sec.~\ref{sec_eq} and their importance with respect
to the concept of objectivity in a final theory, in Sec.~\ref{sec_coms}.
In Sec.~\ref{sec_evidence} we draw first conclusions and discuss their implications,
mainly that the foundation of a final theory can only be formulated as a tautology.
In Sec.~\ref{sec_doing} we describe the tautology, namely we provide a simple 
description of the premises of physics.

In Sec.~\ref{sec_appearance} we scetch some consequences concerning the 
attitude of science towards reality. We shall argue that the formulation of
a \FiT first of all requires to withdraw all metaphysical presuppositions.
In Secs.~\ref{sec_dynamical},\ref{sec_survey} we scetch the conclusions for the
form of a final theory. In Sec.~\ref{sec_realism} we shall argue that 
there is an intrinsic incompatibility of rationalism with realism.

Finally we shall summarize our considerations in Sec.~\ref{sec_summary}.

\section{The specifications of a final theory}
\label{sec_fit}

The current idea of a \GUT is that of a theory which unifies
all {\it forces of nature} into a common conceptional framework. We do not
intend to discuss the various candidates on the market like quantum gravity or
the various string theories. It is sufficient to notice that these theories 
aim to be {\it unified} theories. But a {\it final} theory is more than
just a unification, it is a theory which leaves no questions open. It not
only allows to derive all {\it known} types of particles and interactions,
but also the dimension of space-time. But even this is not sufficient to
provide finality: As we shall argue, the finality implies that it even 
provides a formal derivation of the fundamental concepts of physics. 

This specification might seem exaggerated and the idea of the possibility of
a \FiT hubristic, however this is not necessarily the case. The idea of a \FiT 
only exemplifies the ratio behind the scientific project itself: Weinberg's 
dream is a consequence of the well-known ``principle of sufficient reason'' 
(PSR) and it claims nothing more than that everything in nature should have a
reason~\cite{Reason}. Indeed it is difficult to imagine and to argue that some 
fact, or law, or relation in nature might have no reason at all, that
it might be based on pure contingency, that the mass ratio of proton and 
electron is unexplainable. It would simply be unreasonable and alien to
science to claim that physical relations or some law might have no reason.
Because - if {\it something} in nature has no reason, why should 
{\it anything} have reason? Hence science has to presume reason, and that 
is why it is reasonable to consider that a \FiT could be within reach some 
day - since anything that is reasonable can, by definition, be understood 
~\footnote{This does not imply that everything must have an intelligible 
cause, though.}.
Technical or financial limitations might prevent physics from performing 
the required experiments, or the mathematics might become so difficult, that 
we will never be able to solve all riddles of physics. But to question the PSR 
would go beyond that - to some degree it would mean to question the scientific
project itself.

Gerard t'Hooft expressed his attitude towards the idea of a \FiT as follows~\cite{ToE2005}:
``[...] in particular string theorists expect that the ultimate
laws of physics will contain a kind of logic that is even more
mysterious and alien than that of quantum mechanics. I, however, will only be 
content if the logic is found to be completely straightforward.''
Or Leonhard Susskind (ibd.): ``Not only would the theory explain why the proton and 
neutron are about 1,800 times heavier than the electron, but the theory would
also explain itself: no other theory is possible.''
If we include these requirements in the specifications of a \FiT, then 
the theory has to carry the evidence for its truth in itself. This means that a 
\FiT has to demonstrate that the laws governing {\it physical reality} 
could by no means be different than those formulated by the \FiT. 
Then of course the \FiT has to provide a formal certainty and 
``logical rigidity'' as otherwise only known in mathematics. 

Stanley Goldberg characterized the logic of mathematics as follows~\cite{Goldberg}:
``Different branches of mathematics have different rules but in all branches, 
since the rules are predetermined, the conclusion is actually a restatement, 
in a new form, of the premises. Mathematics, like all formal logic, is
tautological. That is not to say that it is uninteresting or that it doesn't 
contain many surprises.`` 
Hence a \FiT might be possible if it is formally equivalent to mathematics.
This raises the question of the differences between mathematics and physics 
and if and how they could be overcome. We shall come back to this in 
Sec.~\ref{sec_math}.

The required self-evidence of a final theory (considered that such a theory
is possible) does not allow to base the theory on substantial
propositions, since substantial propositions are not self-evident.
Using Goldbergs phrase we might say that substantial propositions are 
those that are not predetermined.
All propositions that can be derived within the \FiT must be inherent 
in some form in the premises of physics or they must be derivable 
mathematically. To substantiate the meaning of this idea, let us consider
some examples of what propositions might be regarded as substantial. 
It is a substantial claim that space-time must have $3+1$ dimensions
and to believe that space is fundamental.
Though it is clear that our world {\it appears} to have this form, it is 
not evident why this {\it should} be the case - why {\it any} physical world
should have $3+1$ dimensions. One might have doubts that physics will 
ever be able to give clear and distinct reasons why any possible physical 
world must have $3+1$ dimensions, but one can not deny that the question 
``why 3+1 dimensions?'' itself is legitimate and meaningful. Therefore a 
statement that the physical world has (or must have) 3+1 dimensions, is 
substantial and requires a derivation within a \FiT. 
Hence, if we believe that a \FiT should be possible and that it has to 
provide such explanations, then {\it any} physical world {\it must} 
have $3+1$ dimensions. 

A \GUT might be based on substantial assumptions and postulates, for instance 
a specific dimensionality of space-time or some special general group 
structure or on the postulation of some type of string: A \GUT could be 
formulated as a theory of {\it postulates}. However, Einstein remarked that 
``When we say that we understand a group of natural phenomena, we mean that 
we have found a constructive theory which embraces them''~\cite{Einstein20}. 
Following Einstein, a \GUT might as well be a theory of principle, but a
\FiT has to be anticipated as a purely constructive theory: It has to explain 
the $3+1$ dimensional space-time by construction, it has to explain, why 
and what physical significance specific groups have, why there must be a 
maximal velocity of massive objects, why Maxwell's equations have their 
specific form and so on. This implies that a \FiT (if possible at all) 
also has to reformulate relativity; it has to replace all of its postulates 
by a reasonable construction. The same would hold for quantum mechanics.
A final theory has to explain why we (have to?) use complex numbers in
quantum mechanics and why the momentum should be equal to the spatial 
derivative of the wave function (times the unit imaginary). A \FiT is 
a deductive theory in which all elements are deduced from unquestionable 
first trivial principles.

\section{What distinguishes physics from mathematics?}
\label{sec_math}

Since the final theory is constructive, it can not simply presume concepts 
like mass or velocity that originate from experience. The required 
inevitability implies that a final theory has to demonstrate how quantities 
like mass, energy, charge and field emerge within a mathematical framework 
that is developed on the basis of self-evident abstract principles. 
However there is at least one concept that has to be presumed: Time. 
The reason (and legitimization) is the
inevitability of this concept. There is not way to describe the meaning of 
physics without presuming the notion of time in some way.

Goldberg described this in his book on relativity~\cite{Goldberg} as follows:
``The undefined terms for the dimensional system we will be using are 
"length," "time" and "mass." These are the common, undefined primitives of 
physics. Although I can tell you how to measure a length, or a time, 
or a mass by a formula, the concepts themselves are undefined. Either
you know or you don't know what I mean when I use a phrase like "time passes."
The dimensions of all other quantities in physics can be expressed in 
terms of these fundamental quantities.'' 
This means that the ``understanding'' of certain concepts can only be obtained
by personal experience. If all substantial concepts of physics have to
emerge within the final theory then a \FiT is formally equivalent to
mathematics. And then the ``mapping'' between mathematical form and physical
meaning is reversed with respect to the usual method of physics. 
If substantial physical concepts like ``mass'' and ``energy'' or ``spin'' can 
not be presumed, these notions have to be identified from pure mathematical
form. 

Usually the arrow that describes the formation of concept in physics, points from the
phenomena to the equations. In mathematics one also finds the opposite
direction: Abstract concepts that have been invented on the basis of pure
formal considerations and then obtain a name from conventional language like
``fiber'' or ``pencil''. 
To establish a relationship between the mathematical form and its physical
content is called {\it interpretation}. And in contrast to mathematics, the 
interpretation charges variables with {\it physical meaning}: the math
of physics refers to things outside mathematics. It would be meaningless to 
ask whether names given to purely mathematical objects or functions 
are ``correct''. Mathematical objects do not refer to anything else outside
mathematics. But a \FiT has to identify mathematical structures with 
physical objects. There is no formal way to prove the truth of this mapping.
We can only verify (or falsify) its experimental predictions. But the
variables that correspond to these predictions are a subset of all variables
of the model: We measure {\it mean} values - even in the microscopic case.
Phenomenological physical models that describe specific physical problems
are to a large degree self-sufficient. It is their purpose to provide 
predictions. But a \FiT has a different purpose, namely to provide the
bridge between phenomenological theories. It must be complete 
and convincing, though it is questionable, whether the required 
interpretations can be unique: the effectiveness of physical models is 
that they are applicable to a variety of phenomena.

If the phenomena (the meaning) has not been established within physics 
beforehand or if the \FiT would describe completely unknown phenomena
of nature, then the attempt of interpretation might be a serious challenge:
The {\it meaning} of variables can (safely) only be obtained from already 
known physical concepts. That is, the path of reasoning within a \FiT can 
not start at the phenomena and end in equations, but vice versa. 
The equations have to be derived on the basis of the abstract principles and
the resulting variables in these equations have to be identified with 
physical quantities. This is a task of {\it pattern recognition} or - as 
Walter Smilga named it - ``reverse engineering''~\cite{Smilga}.

This is a characteristic difference between mathematics and the classical
method of physics: Mathematicians define or derive abstract objects which are
invented in a purely formal way, as a kind of game, a game of thought. 
Then, in a second step, the forms obtain their names, which are essentially 
arbitrary. Physics however picks properties known from everyday experience 
on the basis of common language. These properties have to be 
defined scientifically but this should not alter their meaning, it should
just render their meaning scientifically, i.e. quantitatively. 
The meaning of the ``mass'' of a material body is essentially the same in 
physics and in common sense, or at least it is very close. 

When the physical significance of some symmetries have been understood,
then physical theory might predict new particles because the specific symmetry
principle suggests the particle for its completeness. But to invent a new
symmetry principle {\it without} any prior experimental knowledge can be
regarded as almost impossible. Hence the \FiT {\it requires} for its formulation
that most pieces of the final puzzle are already known~\footnote{
Neither Leibniz nor Platon nor Newton could have derived a final theory from
pure thought, simply because the {\it phenomena} to be explained where not 
known precisely enough. Platons list of solids insofar proves the impressive 
power of pure thought as Platon obviously understood the importance of 
symmetry principles. But Platon could not have arrived by any means at the
gauge group $SU(3)\times SU(2)\times U(1)$, not even if the mathematics would
have been available: Without the knowledge of the phenomena, i.e. the list
of particles, there would be no indication of its physical significance.}.

Elementary particle physics is a special case in this respect and its methods
are closer to those of mathematics: Most properties (or components) of particles 
are abstract and have no counterpart in the macroscopic world, like for
instance (iso-) spin or parity or helicity. In this case physicists have 
to {\it invent} new, essentially arbitrary, names for particle substructures 
(``quarks'') and properties (``strangeness'', ``charme''). The deeper physics 
dives into the micro-cosmos, the closer it seems to be to pure mathematics.
Without the knowledge of the empirically developed physical concepts and
notions it is hardly possible to recognize the physical meaning of 
the variables appearing in the supposed final theory and to find interpretations 
that provide a physically {\it meaningful} theory from purely formal (mathematical) 
relations. If a \FiT is able to deduce equations from abstract principles, then
the ``correct'' interpretation is not included for free.

Physics and mathematics require certain notational conventions, a nomenclature, 
that assigns symbols to certain quantities and relations. 
For instance, $a^2+b^2=c^2$ is associated with the Pythagorean theorem, 
while $m^2+p^2=E^2$, though of identical form, reminds us of the relativistic 
energy formula in units where $c=1$. 
A purely formal derivation of equations within a \FiT as suggested by an 
analogy with pure mathematics does neither provide the correct symbols nor 
the correct interpretation: it only provides mathematical relationships, the 
form. If concepts like ``mass'' and ``energy'' are supposed to emerge from the 
formalism then, at some stage, the \FiT must provide an equation that 
matches, for instance, the mentioned {\it form} of the relativistic energy 
formula. However the \FiT can not provide a proof that the derived formula really 
{\it means} the relativistic energy formula. A formally derived \FiT does
not automatically provide {\it physical} but rather {\it mathematical}
relationships. The form provides information only in the eye of the well-prepared 
observer. The transformation of a mathematical form into a physical meaningful 
law requires an identification of the terms appearing in equations with physical 
quantities. This process is to some degree heuristic: It is an interpretation. 
Therefore it is unlikely that equations that emerge within a final theory can 
be interpreted correctly by pure thought. The formulation of the correct 
principles alone is not sufficient to obtain physical meaning: 
Physical laws have to be known and proven meaningful in advance, before a
\FiT can be formulated. Modifying Goldberg's insight, we might say that we 
already have to know the form and relevance of the Lorentz group in order 
to recognize its significance within the mathematical framework of a \FiT. 
Therefore it is unlikely that a \FiT could be formulated ``from pure thought'' 
before physics sufficiently progressed towards a fundamental level of physical
reality: We can only recognize known patterns; the pure form does not generate
meaning. When the abstract inevitable principles of a \FiT are found,
and a mathematical formalism has been developed then one still has 
to establish the mapping between an abstract mathematical form and known 
physical laws. This step in the formulation of the \FiT might be summarized 
as ``function follows form''.
Though the principles of the \FiT must be trivial (``self-evident''), this
does not necessarily hold for the formulation of the theory. Therefore it
seems unlikely that a \FiT will - from beginning on - be recognized as
a possible candidate for a theory of everything. This seems to be sufficiently
proven by contemporary candidates like string theory: The initial ideas of 
theories might be simple, but the resulting structures and mathematical
forms must (by definition) be rich and complicated enough to allow for the
description of a large number of particles and fields. But if they are
rich and complicated enough, then they are difficult to interpret.

The basic system of ``undefined'' quantities, that Goldberg mentioned,
is still debated, but it appears obvious that certain quantities can 
not be defined by other means than by a description of their respective 
measurement method or directly by reference to experience. This is 
for instance the case for the notion of time. We can not ``explain'' 
time to someone who does not know what it is because it is {\it unique}. 
A unique concept can not be explained by other similar concepts since 
it is unique. But though clocks don't explain or define what time {\it is},
clock's define how we {\it measure} time. And this is all that is required:
Physics is neither able nor obliged to tell {\it what} things are,
but physics can tell us {\it how} things behave. It could be argued that,
from the complete knowledge of how things behave, we finally obtain an
idea of what things are. At least one might finally say, ``A'' and ``B''
behave {\it as if} they were ``X'' and ``Y''. The quantum mechanical
wavefunction behaves {\it as if} it was a probability density, but
we can't tell seriously that it really is (and nothing else). And it is not
even obvious what difference it makes.

The impossibility to define the concept of time is insofar unproblematic 
in this context as the concept of physics itself presupposes time: The 
formulation of the ``laws of nature'' is meaningful because laws allow 
to predict measurement results. Physics is and always was more than an 
intellectual excercise. It is a science that aims to provide new 
technologies, new methods that allow to manipulate objects. 

A physical prediction is in several ways related to time: First of all,
predictions enables to know the value of a measurable
quantity {\it before} the measurement has been done, or, as in quantum
mechanics, it at least enables to predict the possible measurement 
results and their probabilities. And secondly there is an indirect 
relationship by the presumptions that are required to make a measurement 
meaningful, namely, that we have reference objects with {\it constant} 
properties (a clock, the prototype meter, etc) and variable properties 
that are to be predicted. No measurement without a reference. 
Hence the form of physics implies and presupposes that there is time, 
i.e. that the same type of quantity is allowed to appear in two forms, 
as a variable and as a constant reference.
 
Therefore we shall next discuss the role of units and of so-called 
``physical constants'' in Sec.~\ref{sec_ref}. 
Inherent in the notion of time is also the concept of constancy and
the specific role of constants is an important difference between 
physics and mathematics. Some of these constants are called 
``physical constants'' {\it because} they do not appear in mathematics. 
These are first of all the constants with dimension, i.e. those that 
have units, like the speed of light $c$ or the Planck's 
constant $\hbar$. Then there are dimensionless constants like the
fine-structure constant. The latter could, at least in principle, be the 
result of mathematical relations. A final theory has to provide the 
values of these dimensionless constants - which implies that they 
can be derived.

\subsection{Measurement Standards And Physical Constants}
\label{sec_ref}

Physics used arbitrary measurement standards in its history. 
There is the tale about Galileo, who measured the frequency of swinging
oil lamps in church using his pulse as frequency standard~\cite{Modinos}. 
The use of the own body as a basic standard is a natural and obvious
thing to do, as the size of an object relative to ourselves is exactly what 
we naturally want to know. The number of steps on the way between 
point {\it A} to point {\it B} is exactly the type of practical knowledge
that one may assume as the origin of measurement. And quite naturally, 
distances can be and likely have always been measured in units of time: 
{\it A} to point {\it B} might be a three day walk or a single day horse
ride. The use of the own body as a reference to measure length is nearby, 
because the scale of our body defines the scale of our world. 

It is well within proportion that the meter is roughly the size of a
footstep. The unit of a {\it foot} testifies this finding and 
shows that measurements started with scientifically arbitrary units - 
standards that are defined pragmatically. 
The first modern scientists like Galileo picked phenomenological 
properties like length or duration of something in motion and measured 
dynamical properties of swinging pendula and falling stones. Consequently the 
first ``laws'' are mere (idealized) descriptions of proportions~\cite{Boccaletti}.  
First scientific observations were of the sort, that the frequency of a 
swinging pendulum does (in first approximation) not depend on its 
amplitude. This continued with Keplers and Newton's laws: Physical laws 
were rarely formulated as equations, but as statements of
proportionality. It should be noted that {\it laws of proportionality} 
allow to circumvent the definition of universal units. If we say that the frequency 
of a swinging pendulum is inversely proportional to the square root of its length, 
then the experimental confirmation requires (if at all) rulers and clocks.
The law itself is correct no matter what units are choosen and hence a 
universal system of units is not required to confirm them experimentally.
Despite the fact that it is possible to formulate {\it objectively} correct 
laws of proportion without explicit reference to specific objects as
measurement standards, the measurements that are used to find these laws, 
of course require rulers and clocks. But they do not need a universal gauge. 

As technology evolved it became unavoidable to define systems of units: laws 
of proportionality do not suffice for the construction of a pendulum with 
a specified period. A consequence of the formulation of the laws of nature by 
the use of equations instead of mere laws of proportions is the appearance 
of {\it conversion factors} like the Avogadro constant or the gravitational constant. 
Often these constants are named {\it physical constants}. Wikipedia explains
the term {\it physical constant} as follows: ``A fundamental physical constant 
is a physical quantity that is generally believed to be both universal 
in nature and constant in time.''~\cite{Wiki}. This, however, does not reflect
the understanding of the physical significance of physical constants in
theoretical physics. Dirac wrote: ``The information that 
experimentalists obtain provides us with a number of constants.
These constants usually have dimensions, and then of course, they depend 
on what units one uses, whether centimetres or inches. Then they are not 
of theoretical interest.''~\cite{DiracLNH}. Dimensionful physical constants 
bear no physical significance. This point of view has not changed
significantly until today~\cite{Hsu} - even the idea that Planck's constant 
is physically significant was questioned with some convincing arguments~\cite{RalstonHBAR}. 

This situation might appear paradoxical: Those physical constants 
that can expected to be the result of purely mathematical considerations, 
are supposed to be those with physical significance and a \FiT would eventually
have to provide their values, while those with a physical dimension, 
are considered to be physically meaningless. 
But though the numerical {\it values} of dimensionful physical constants are 
meaningless, as Dirac explained, their {\it constancy} is not. For instance, 
the numerical value of the speed of light is meaningless as it depends on the 
choice of units. But the fact that a {\it maximal velocity} for massive objects 
exists, {\it is} of course meaningful, it is even of fundamental importance. 
However, since a \FiT must be formally derived from abstract principles, 
the values of dimensionful constants can not be the result of such a theory.
And since their values are not significant, there is no reason why these 
constants should emerge from fundamental equations. In theoretical physics, 
the system of units is often chosen in a way that the dimensionful constants 
like $c$ and $\hbar$ can be replaced by unity, i.e. they do not appear in the
equations and we can expect the same for a \FiT.

For instance, it can be shown that the group velocity $v_{gr}$ of 
wave-packets is given by $v_{gr}=\nabla_k\,\omega({\bf k})$. 
In quantum mechanics (QM) and quantum field theory (QFT), wave packets 
represent particles and therefore the velocity of these particles
is derived from the respective dispersion relation. For massless particles
this yields $\omega({\bf k})=\vert {\bf k}\vert$ such that $\vert v_{gr}\vert=1$. 
If frequency and wave vector have the same unit, there is no theoretical
need for a dimensional constant called ``speed of light''. 
This means that a \FiT constructed from formal principles can not
provide equations in which the so-called ``physical constants''
$c$, $\hbar$, $\eps_0$ automatically appear. In a \FiT such constants have to
be introduced {\it artificially} simply because these constants {\it are
artificial}.

\subsection{The Significance of Natural Units}
\label{sec_natural_units}

Dimensional physical constants loose their contingent status only if 
{\it natural units} like the elementary charge $e$ or the electron mass $m_e$ 
can be found which allow to replace the contingent historical units with 
``objective'' units, i.e. the corresponding constant properties of 
fundamental objects~\footnote{
We emphasize the importance of reference standards within the logic 
of physics: objectivity is based on the use of {\it reference objects}. 
Let us expand the notion of objects and include properties of 
{\it reference processes}, despite the fact that we will later conjecture 
that {\it objects} can - within a final theory - only be understood as 
processes.}.
One of the first known natural units was the {\it elementary charge} $e$ 
and it is indeed remarkable that any amount of charge in nature can be 
expressed by an integer number. If quarks are considered, the natural unit
charge is $e/3$, but still any natural amount of charge is quantized and 
can be expressed as in the natural unit represented by some fundamental 
object that serves as reference. 

{\it Natural units} refer to the properties of fundamental (stable) objects 
(or processes) of nature. If matter would not be composed of
atoms, e.g. if matter would not be quantized in {\it some} way, such 
fundamental objects could not be explained within a \FiT. It was a common 
idea in the history of physics that matter would be something that completely 
and continuously fills up space, pretty
much as it {\it appears} to the eye. This idea is nearby, directly derived 
from the ``empirical evidence''. Today we know that the nucleus is very small
compared to the atom's dimensions, which are defined by the electronic orbitals,
so that matter is far from filling up space. 

However, if matter would really be a continuous {\it something} that fills 
up space-time in a way that could not be derived from a more fundamental 
theory, then a \FiT would be impossible, because this would imply that
space-time is fundamental and hence unexplainable. It has been suggested 
that the antropic principle might serve as an explanation. But it hardly
{\it explains} anything; it does not generate the kind of reason that
the principle of sufficient reason recomments.

Without a description of the internal dynamics of such a material, 
we could by no means argue that and why a specific radius should appear.
The only way to obtain a granular structure of matter would then be
to {\it postulate} some kind of quantization. Otherwise there would be no 
means to replace the arbitrary units of mass and charge by natural 
units and a final theory in the sense of Weinberg would be impossible. 
Without quantization we could not refer to the respective properties of an 
electron and it seems clear, that a \FiT would be unthinkable in such a 
world as there would be no way to get rid of the arbitrariness of 
dimensionful units. This implies that a classical \FiT (i.e. a \FiT without
some type of quantization) is logically impossible. And vice versa: If we 
speculate that the formulation of a final theory is possible, then the \FiT
has to contain quantization in some way. Otherwise we could not derive
the objects that provide the required natural units.

\subsection{The Form of Physical Equations}
\label{sec_eq}
 
Given two physical quantities $E$ and $p$ of different dimension are
supposed to be related by a physical law in the form
\begeq
E=f(p)\,.
\label{eq_EfromP}
\endeq
Maybe we derived this law by a sequence of measurements $E_k$
in dependence of $p_k$. The range within which one can experimentally prepare 
$p$ and measure $E$ is certainly finite, such that one can always approximate 
$f(p)$ by a polynomial of finite degree. 
In the simplest case, we find that $f(p)$ is proportional to $p$ or $p^2$.
If $f(p)$ is a steady function, this has to be expected for small enough
ranges of $p$, as one may always write a continuous function $f(p)$ of a 
continuous variable $p$ as a Taylor series:
\begary{rcl}
f(p)&=&\sum\limits_{k=0}^\infty\,\left.{df^{(k)}(p)\over
  dp}\right\vert_{\tilde p}\,{(p-\tilde p)^k\over k!}\\
    &=&C_0+C_1\,(p-\tilde p)+C_2\,(p-\tilde p)^2+\cdots\\
\label{eq_taylor0}
\endary
However, we can only add physical quantities of the same unit. Since $p$ 
has a dimension, then all coefficients $C_k$ have different dimensions, 
$C_0$ the dimension of $E$, denoted by $[E]$, $C_1$ has dimension $[E]/[p]$, 
$C_2$ the dimension $[E]/[p^2]$ and so on. 
This would lead to an infinite number of dimensional constants.
However, as only two dimensionful quantities are related in
Eq.~\ref{eq_EfromP}, it is sufficient to establish two dimensionful
non-zero constants $p_0$ and $E_0$ and to express $f$ in the form
\begeq
E=E_0\,f(p/p_0)\,,
\label{eq_EfromPnormalized}
\endeq
where now $f(x)$ can have any mathematical form since $x=p/p_0$ is
dimensionless. These two dimensionful constants would fall into one,
if the Taylor series had just a single term, i.e. if the true law
would be 
\begeq
E=E_0\,(p/p_0)^n\,,
\endeq
where we assume $n\ne 0$ to obtain a non-trivial equation. However,
in this case, we could not determine two constants $p_0$ and $E_0$
separately, but only the proportionality constants $C=E_0/p_0^n$.

This example demonstrates that the mathematical form of physical laws and the 
number of physical constants are indeed related. If a physical relation
is continuous and can be written as a Taylor series (Eq.~\ref{eq_taylor0})
{\it and} the required number of natural units $E_0$ and $p_0$ can
be found, then one can rewrite Eq.~\ref{eq_taylor0} in the form
\begary{rcl}
x&=&p/p_0\\
E(x)&=&E_0\,\sum\limits_{k=0}^\infty\,\left.{dE^{(k)}(x)\over
  dx}\right\vert_{\tilde p}\,{(x-\tilde x)^k\over k!}\\
    &=&E_0\,(1+D_1\,(x-\tilde x)+D_2\,(x-\tilde x)^2+\cdots)\\
\label{eq_taylor1}
\endary
with an (potentially) infinite number of dimensionless constants $D_k$. 

If the values of these constants would be independent such that we
could not find a mathematical law that allows for their derivation, then 
such a law needed an infinite number of dimensionless physical constants
that would have to be the result of a \FiT. But since the \FiT must itself
be analogue to mathematics, this would be a contradiction in itself.
As we specified that a \FiT must derive all dimensionless physical 
constants {\it mathematically}, then the most fundamental laws may not contain
meaningful dimensionless physical constants.
The idea of a \FiT implies that the dimensionless physical constants are 
{\it effectively} mathematical constants. Hence the Taylor series is a 
short polynomial or it is a transcendental function. 
Since the most fundamental equations of a \FiT may not include dimensionless 
constants, these most fundamental laws have to be simple. Not because 
physicist prefer simplicity, but because the laws are assumed to be 
fundamental. Nature might turn out to be different, but if it is, then 
the specified \FiT is logically impossible.

Let us assume the true law would be
\begeq
f(p)=E_0\,(\exp{(p/p_0)}-1)\,,
\label{eq_true}
\endeq
If $p_0$ is very large compared to the experimentally realizeable range of
$p$, then the first experimentally verified law might be
\begeq
f(p)=C_1\,p\,.
\endeq
This law would neither be absolutely right nor absolutely wrong. It would
be adequate within a certain range of $p$ and increasingly inaccurate and finally 
useless, if the $p$ of interest is not within or not even close to the
mentioned range. It is possible to determine $C_1$ by the measurements within a 
limited range of $p$, but not $E_0$ and $p_0$ separately. With progress 
of the experimental technology one might be able to extend the range of $p$ 
towards and beyond $p_0$, such that we would be able to find
\begeq
f(p)=E_0\,\left({p\over p_0}+\frac{1}{2}\,\left({p\over p_0}\right)^2\right)\,,
\endeq
and maybe with further experiments (or with a deeper theoretical insight)
we might conclude that the true law would have to be Eq.~\ref{eq_true}.
It is obvious that the first dimensionful constant $C_1$ has, as 
such, no fundamental meaning. But what about $E_0$ and $p_0$?

Such constants have fundamental physical significance given that 
they represent {\it natural} units $E_0$ and $p_0$ {\it or} that they 
can be expressed by other natural constants. 
In the relativistic energy-momentum relation 
\begeq
E=\sqrt{m^2\,c^4+p^2\,c^2}
\label{eq_EofP}
\endeq
these two constants are $m$ and $c$, such that with $E_0=m\,c^2$ and
$p_0=m\,c$ one may write instead of Eq.~\ref{eq_EofP}
\begeq
E/E_0=\sqrt{1+(p/p_0)^2}\,.
\endeq
The mass $m$ is then a property of the electron (for instance).

As a final theory may not contain arbitrary units but only natural units
as references, these units play a fundamental role for the way, in which 
physical laws are and can be formulated within a \FiT. The natural units 
however are properties of {\it reference objects}. The \FiT can not derive
the {\it value} of the mass of a reference object: The \FiT expresses the
mass of other objects in units of the reference object.
The \FiT can also not explain the stability of the reference object: The
\FiT instead uses the stability of the object as a self-sufficient fact:
In an objective reality we must have stable objects.

However, the \FiT may not presume dimensionless constants since the
{\it derivation} of dimensionless constants was one of the requirements 
for a final theory. The conclusion of these considerations is that a \FiT 
can not refer to {\it any} physical constant at the most fundamental level. 
Nevertheless it is supposed to provide {\it objects} with constant
properties which may then serve as references and that provide natural units. 
This apparent paradox can (only) be solved, if the constant properties of 
the objects are constants of motion (COMs) as we shall explain in 
Sec.~\ref{sec_survey}.

Only the existence of fundamental objects allows to refer to fundamental 
quantities. This in turn allows to express physical quantities by real 
(or even by natural) numbers. The final theory became thinkable only
by the discovery of the sufficient number of physical constants, e.g. 
by the invention of relativity and quantum mechanics. Without the natural 
objective constants $c$ and $\hbar$, a final theory can not be formulated 
due to the lack of reference quantities. Again: The values of these constants 
are not meaningful, but their constancy is. 
This is the modern understanding of physical units and constants: 
A final theory contains no arbitrary units, because every quantity can be 
expressed by the corresponding property of some fundamental object of nature. 

Dimensionless constants like the fine-structure constant 
$\alpha\approx 1/137$ or the mass ratio of electron and protron 
$m_e/m_p\approx 1/1836$ represent proportions between the properties 
of fundamental objects or processes. A \FiT should provide the 
mathematical arguments that allow to derive these proportions and in case
of success these constants are effectively {\it mathematical constants} 
like $\pi$, which represents an extremal proportion between circumference 
and diameter.

\subsection{Reference Objects and Constants of Motion}
\label{sec_coms}

On the most fundamental level, a \FiT may neither refer to dimensional constants
nor to dimensionless constants, because the former are arbitrary and the latter
have to be derived within the \FiT. But on the other hand objects with constant 
properties are required as references in order to allow for measurements. 
This apparent contradiction can be resolved if we draw the following conclusions: 
Firstly, all quantities that belong to the fundamental level, must be 
{\it variables} in the sense that they have to vary at all times. 
Secondly, since constants are required to enable for measurements, then the most 
fundamental level obviously does not allow for a direct measurement. 
However, there must be a one or several functions of the fundamental variables 
that generate constants of motion and therefore the reference constants must 
be constants of motion. If all fundamental variables continuously vary, 
the constants of motion must at least be of second order in terms of the 
fundamental variables. 

Constants of motion appear in all dynamical laws of physics. The best known
constants of motion are energy, momentum and angular momentum and a \FiT
has to provide a mechanism from which these COMs emerge. From Noether's
theorems we know that constants of motion and invariance properties are
closely connected. As described in Ref.~\cite{qed_paper,osc_paper} the 
suggested trivial principles allow to derive Hamiltons equations of motion 
(EQOM). But the fundamental variables are not ``coordinates of mass points'', 
but purely abstract variables that can not be directly measured. This is a 
logical implication of the idea of a \FiT and if we interpret fundamental 
variables as (components) of quantum mechanics wave functions, then we found
a reason why the wave function itself can not be directly measurered:
Quantum mechanics has the task to analyze the behavior of ensembles solely
on the basis of their observables. But the fundamental objects, represented
by the wave function, can not be directly measured: There is no direct unit
for the wave function, because the (components of the) wave function 
{\it vary at all times}.

A consequence of this argument is that truely fundamental quantities 
{\it can never be subject of a direct measurement}.
This is the trivial and evident reason, why we can not measure (components
of) $\psi$ and claim that ``the first component of the wave function at 
position ${\bf x}$ is five pumis''~\footnote{A pumi is a fictious
unit for wave functions.}, while there is no (principle) problem to obtain 
the value of a magnetic field $B({\bf x})$ at some position in units of 
Gauss or Tesla. Then, however, it follows that a magnetic field value 
can not be regarded as a {\it fundamental} quantity: no directly measureable 
physical quantity can be regarded as fundamental.

\section{What could provide the trivial evidence of a final theory?}
\label{sec_evidence}

If we apply Goldbergs characterization of mathematics to physics, then a
\FiT has to be a tautology. As mentioned in the introduction,
we suspect that a \FiT might be formulated on the basis of what we {\it do}
when we perform physical experiments, physical measurements, i.e. what we
{\it have to} assume unless our experiments and measurements are meaningless.
Here we might also find the final answer to the question in what respect 
physics differs from pure mathematics. 

Therefore the first step towards a \FiT would be to provide an inventory 
of the primary principles that are trivially true in any world that allows 
for the formulation of physical laws and to remove all presumptions that 
can not be considered to be trivially fulfilled. Since {\it substantial} 
presumptions can not be preconditions for the formulation of a \FiT,
they should be critically reviewed. Essentially this requires a 
description of the {\it form of physics}.

\subsection{The Form of Physics: What Physicists {\it Do}}
\label{sec_doing}

Mathematics is itself not a model of anything but a logical framework,
and also a \FiT can not {\it in it's origin} be a model of anything.
Hence it can not be directly derived from the phenomena.
If a \FiT is possible, then the description of the physical aspects of 
reality is based on a set of formal principles and we will show that 
the form of physics provides trivial principles that can be used as
a starting point.

Physics is a scientific discipline because it is based on a well-defined 
method to test theories. The method of physics is ``objective'' - but not
because it describes ``real'' objects or some kind of ``objective reality''.
It is objective because it is based on the comparison of objects with
objects. Measurements are comparisons - physicists compare the size, 
weight, velocity, energy and other quantitative properties with the size, 
weight, velocity and energy of {\it reference objects}: 
the prototype meter, the kilogram and the velocity of light are - or have 
been - used as reference.
The branch of physics that provides the required references, is called
metrology. The definition of units and the {\it production} of reference 
artifacts might appear to be an {\it inevitable} but merely formal act. 
However, it is exactly of the kind of {\it inevitability} which is required 
for the formulation of a \FiT. The inevitable aspects of physics are 
those which in summary are the form or method of physics. 
Since physical theories are tested by measurements, the things that 
physisists {\it do} when they perform measurements, is an essential 
part of the {\it form of physics}.

Einstein wrote about his theory of special relativity that
``It is striking that the theory (except for the four-dimensional space)
introduces two kinds of things, i.e. (1) measuring rods and clocks, (2)
all other things, e.g., the electromagnetic field, the material point, etc.
This, in a certain sense, is inconsistent; strictly speaking, measuring rods
and clocks should emerge as solutions of the basic equations [...], not, as
it were, as theoretically self-sufficient entities.''~\cite{EinsteinBio}.
This confession should be surprizing and disconcerting to physicists 
and philosophers of science, as it suggests that there might be missing piece, 
a systematic incompleteness in the logic of objectivity of the natural
sciences. However (to the knowledge of the author), the fundamental
role that metrology plays with respect to the method of physics is rarely 
mentioned, discussed or even noticed, despite the fact that the use 
of objects as reference standards is undeniable the fundament of 
objectivity in physics. A final theory, if considered possible at all, 
has to incorporate the trivial and fundamental role played by measurement 
standards.

In the mentioned quote, Einstein describes what he did when he formulated 
special relativity, but it is just an example of what physicists always do, 
namely to introduce references combined with certain invariance assumptions. 
As they seem blatantly trivial, these assumptions are rarely spelled out 
explicitely; but to define rods and clocks (e.g. some objects) as 
references, implies a number of presumptions about the relevant properties 
of these objects, namely that they are {\it invariant}.
First of all they are assumed to be constant in time, but the meter rod
is also assumed to be invariant with respect to its orientation in space.
Certainly we won't suggest to doubt these invariance principles, but we 
should be aware of the implications of this method. Because the presumed 
invariance principle leads, if violated, to the introduction of fields: 
If the spin of an electron has no preferred direction, then this is 
{\it phenomenologically} identical to the absence of a magnetic field. 
And vice versa: The violation of the isotropy of space is 
{\it phenomenologically} identical with the presence of a magnetic field.
Contrary to what one might expect, but the presumed isotropy of space 
does not require an experimental proof: 
Where we find space to be anisotropic, we (have to) presume a reason,
namely a (gauge) field. Newton's axiom of inertia obeys the same logic:
According to this axiom a body moves with constant velocity on a straight 
line unless a force is applied. 
Not only that, at the time of Newton, it was technically impossible to 
provide experimental evidence for the presumed straight motion, it is 
logically impossible to prove Newton's premise as it is itself the 
definition of (the absence of) a force. Therefore it is introduced as 
an axiom. 
The underlying principle of this {\it method} is the principle of 
(sufficient) reason. Science has to provide reasons for 
distinctions, for asymmetries, for differences. And if such reasons are 
not at hand, then we (have to) assume symmetry, homogeneity, equivalence.
This is not a principle of nature, it is a principle of thought.

It is one of the revolutionary but also one of the most questionable
aspects of relativity, that it does not simply continue to presume 
the {\it invariance} of rods and clocks, but describes precise conditions 
of invariance and covariance and the corresponding transformation properties.
And in doing so, it apparently breaks with a rule of thought: Rulers are, 
as we explained above, by definition constant.
However, this holds only, if the we assume that space-time is a fundamental
notion, an assumption that is, as we argued above, incompatible with the
logic of a \FiT: it would deny the possibility to argue why space should
be three-dimensional. Whence there is no contradiction between relativity
and the rules of thought, if a \FiT is possible. The \FiT transforms the
contradiction into an obligation: The \FiT has to formulate a criterion 
that allows to characterize dimensionalities and to show how spatial
notions emerge.

Prior to the invention of relativity, physicists simply {\it presumed} 
that clocks and rods are the same for all observers under all conditions. 
The theory of special relativity is the first theory that questioned 
this naive presumption. Now, after the experimental confirmation of
relativity, we believe to know that (perfect) rods are invariant in proper 
time and under rotations, but not under Lorentz boosts. As rods are rarely 
used to measure a length when they move with relativistic speeds, this effect
is irrelevant in all practical situations. It is difficult to imagine a 
situation in which the ``length contraction'' could have been discovered 
by the direct use of rods for length measurements and accordingly it is 
difficult to confirm this ``effect'' by direct measurements. It is a 
consequence of a formalism that has been confirmed by a variety of other 
methods and predictions. Despite this, the theory of special relativity has 
achieved the status of an almost unquestionable corner-stone of physics.
Namely in accelerator and particle physics, it has reached the same status 
of daily applied engineering knowledge as electrodynamics and is usually 
regarded as part of classical physics. To the general public, relativity is 
usually presented as a theory of space-time and rarely regarded as an 
extension of metrology. 
We think that this might be a mistake: Seen by light, Einstein's work is 
full of hints that relativity is intimately connected to problems of 
measurement standards and relative calibration (clock synchronization). 

Our considerations result in the insight that it is not the constancy of 
the length of a {\it specific} (inertial) ruler that we have to presume, but
we {\it have} to presume that regardless of the technical problems to
provide a {\it really} constant ruler, it is {\it in principle} possible. 
This is part of the form of physics: Despite all practical problems, 
we presume that there is no general physical law that prevents us 
{\it in principle} from manufactoring a constant ruler or a perfect kilogram. 
As a final theory has to be free of logical circles, the final theory can 
not presume stability of the natural units and rulers and provide at the 
same time an {\it explanation} for this stability.

A \FiT can not be based on the assumption that a specific property of a 
specific object is constant in time, but it may be based on the more 
abstract assumption that {\it there are constants}. It has to incorporate 
the basic assumption of metrology, namely that nature provides objects 
with constant properties that can be used as references. 
It is then logically impossible to {\it prove} this assumption within the 
theory. Any attempt to do so provides merely the proof of internal consistency: 
It would just show that the theory allows to reproduce the assumptions 
it is based upon.

\subsection{Existence, Appearance and Philosophy}
\label{sec_appearance}

But how at all can it happen that preconceptions based on empirical evidence
fail, that matter is almost empty space, that Newtons concept of absolute
time and space turned out to be wrong? Our preconceptions about space and 
time are taken from ``reality'' as it appears to the senses and as it is 
processed by our mind. But is it not science (and specifically physics) that 
provides evidence and confirmation for the existence of an objective reality? 
Steven Weinberg dedicated a complete chapter of his book to argue 
``against philosophy''~\cite{Weinberg}.
Nevertheless he admits on the second page of this chapter: ``Physicists
do of course carry around with them a working philosophy. For most of us,
it is a rough-and-ready realism, a belief in the objective reality of the
ingredients of our scientific theories.'' 

It is irritating that Weinberg argues against philosophy, as his book 
appears to be more about philosophy than about anything else. But as
he explains, he argues not against philosophy as such but against {\it
  dogmatism}. He writes:
``Although naive mechanism seems safely dead, physics continues to be
troubled by other metaphysical presuppositions, particularly those having
to do with space and time.'' and on the same page he writes: ``Even now,
almost a century after the advent of special relativity, some physicists
still think that there are things that can be said about space and time
on the basis of pure thought.'' 
One might be tempted to remark that the idea of a final theory, as Weinberg 
(pro-) poses it, may as well suit the purpose of arguing in favour
of a theory derived from pure thought: The inevitability, logical necessity
and coherence that he expects from a final theory, are properties 
that can otherwise only be found in mathematics, that is: in a branch of
science that {\it is} based on pure thought. Again we arrive at an irritating
dichotomy. On the one hand, physics defines itself as an experimental science,
but on the other hand it is the final objective of (theoretical) physics
to find a {\it final} theory which must be similar to pure mathematics, in
various aspects. However, the keyword in the above quote is ``metaphysical 
presupposition'' and in this respect the author fully agrees with Weinberg.
A theory should not (and a final theory may not) contain metaphysical
presuppositions. This is a key requirement that we specified for a final
theory: It may not contain any substantial distinctions without sufficient 
reason. Specifically we deny the possibility that a \FiT could be based on 
metaphysical assumptions. Within the context of a \FiT we have to refuse
any assumption that is not based on reason, any distinction that is not 
inevitable. And in consequence this means that we refuse any substantial 
claim.

It appears that Weinberg is aware of the fact that the ``rough-and-ready 
realism'', understood as a habit or a tendency to be philosophically 
attached to the appearance of reality, insofar as it is similar to 
``naive mechanims'', is disproven by experimental evidence. 

In his allegory of the cave, the greek philosopher Plato adressed
the question, whether appearance generates a faithful image of
the true form of reality:
``Plato has Socrates describe a gathering of people who have lived 
chained to the wall of a cave all of their lives, facing a blank wall. 
The people watch shadows projected on the wall from things passing in 
front of a fire behind them, and they begin to give names to these shadows. 
The shadows are as close as the prisoners get to viewing
reality.``~\cite{WikiCave}. In Platos allegory the perception of reality 
does not show something {\it unreal} or an {\it illusion}, but something 
that is a {\it projection} or a {\it derivation} from the true
world~\footnote{Of course, Plato uses the allegory of the cave to claim
that a philosopher is someone who broke the chains and is therefore able
to see the world as it really is. This claim possibly contains a fair 
amount of ``marketing'' for the case of philosophy. Otherwise the famous 
(though slightly adulterated) quote ``I know that I know nothing'' 
would likely not be found in Platos work~\cite{Sokrates}.}. 
Thus our preconceptions fail, {\it because} they are based on notions derived
from the {\it appearance} of the world, namely (as Weinberg emphazises) on wrong
preconceptions about space and time. Our preconceptions fail {\it because}
they are based on ``empirical evidence'', but not on the kind of empirical 
evidence of a physics experiment, but on the kind of naive 
``empirical evidence'' of common sense.

Weinberg wrote: ``It is just that a knowledge of philosophy does not seem 
to be of use to physicists''~\footnote{Chapter 7, ``against philosophy''.}.
Here we clearly disagree with Weinberg. First of all, we believe that some
knowledge of philosophy is useful to any human being, but more than that,
we doubt that the usefulness of philosophy can be derived from the possession 
of some kind of {\it knowledge}. 
Same as physics, philosophy is mainly a proficiency. Because a final theory 
may not be based on metaphysical assumptions, the proficiency to recognize 
assumptions as metaphysical is required.
Regardless on how we arrive at a \FiT - only if the theory can be (re-)
constructed on the basis of pure thought, it can be said to have met his 
specifications. And philosophy (even more than mathematics) is the domain
of pure thought.

Besides the major concepts of physics also the specific (apparently relativistic) 
form of space and time including Lorentz transformations and the
dimensionality of space-time must be the {\it result} of a final theory. 
In order to obtain these notions as results from a final theory, the theory 
has to be based on something else, e.g. {\it not} on the presumption of a 
self-sufficient space-time model. It has to include a part that describes 
the emergence (or appearance) of spatial notions. 

\subsection{The Form of Dynamical Laws}
\label{sec_dynamical}

The example of a physical law as described in Sec.~\ref{sec_eq} concerned 
some general relation between two physical quantities like (for instance) 
energy and momentum.
Such relations allow to reduce the number of free variables by uncovering
a fixed relation between quantities (variables). If we discuss whether a 
specific process (or nature as a whole) is deterministic (i.e. if it
could be predicted by an algorithm), then we refer to a different kind
of physical law, namely to a dynamical law, to so-called {\it equations of 
motion} (EQOM). EQOMs refer to a number of dynamical variables that can
be expressed by the components of a state vector $\psi$ and have the 
archetypical form 
\begeq
\dot\psi=f(\psi)\,,
\label{eq_dyn}
\endeq
- where the dot indicates the time derivative - such that, if the state of a 
system is represented by $\psi$ at a certain point in time, all future 
states of the system can be computed by integrating Eq.~\ref{eq_dyn}. 
We do not consider partial differential equations here, for two reasons: 
Firstly, we said that a \FiT is allowed to be based on the notion of time, 
but not on space - and secondly, 
we expect that the introduction of spatial notions can be delayed to a later 
stage, as it is always possible to use a Fourier transformation to
introduce spatial notions.

In classical physics and namely in classical statistical mechanics, the state 
vector contains the coordinates and momenta (Hamiltonian mechanics) or the 
coordinates and velocities (in Lagrangian mechanics), respectively. 
In the former case, the EQOMs are derived from a conservation law (usually
the conservation of energy) and in the latter case from a variational
principle (the so-called {\it principle of least action}). 
The EQOM depend on a single variable $t$ which can be interpreted as {\it
 time}, the primary ``quantity'' that we accept as being undefinable. 
Time is the (first) variable for which the question ``why time'' makes no
sense within a final physical theory: Already the concept of {\it physical
  variable} and of measurement implies that we refer to change, variation and 
constancy, i.e. to those notions that are associated with time. And we shall 
use the required distinction of variables and constants to discuss the
possible forms of Eq.~\ref{eq_dyn}. 

\section{A Survey of (Trivial) Principles of Physics}
\label{sec_survey}

Our considerations allow to give a first, possibly incomplete, inventory
of formal first principles. The most important being the principle of
sufficient reason. Sir Hamilton summarized this principle in the
short formula: ``Infer nothing without a ground or reason.''~\cite{PSR}.
The idea of a \FiT itself is based on this principle as we explained in
Sec.~\ref{sec_fit}.

The principle of reason (PSR) is the first principle that trivially has to
be respected in a \FiT. The believe that things are comprehensible is a 
precondition for science in general, of course. But the principle includes 
more than that. 
From the specifications of Weinberg, t'Hooft and Susskind it follows that if 
a \FiT is supposed to be inevitable, evidently true, complete and final,
then a \FiT may not contain any arbitary distinction or selection, no 
concept should be introduced without a clear and evident reason. 
The final theory does not have to provide {\it causal} 
explanations for individual phenomena, but it should provide sufficient 
reasons for the general form of the phenomena. Hence a \FiT can not
(and does not have to) tell {\it what time is}, since we can not even
formulate the idea of what physics is without presuming that the concept 
of time is already known.

Secondly, as we explained before, the presumed concepts of time and 
measurement inevitably lead to the trivial confession that physical
models contain variables and constants. As we argued, for logical 
reasons, constants may not appear at the fundamental level, neither 
dimensional nor dimensionless. Despite this, 
the idea of a measurement requires the appearance of true constants
for reference. We might call this requirement ``the principle of objectivity'' 
(POO). It is the basis of the relation between a physical theory and 
``objective reality'': Physical equations refer to relations between 
properties of objects and though this is rarely within the focus of attention, 
these objects have to be {\it physically made}, if the equations are to 
be confirmed. The production of artifacts with constant properties must 
after all be possible in principle. If this is not the case, then we can 
not {\it directly} measure this property. Even worse, strictly speaking 
it could be questioned if such a property is {\it objectively} 
there. 

The dichotomy that we may not directly introduce constants while 
measurements are nevertheless based on the availability of such constants, 
can be solved: dimensional constants are logically connected to the laws of 
motion - they appear in the form of constants of motion (COMs).
Within a \FiT constants of motion are the only possible 
method to introduce dimensional constants. Either we derive 
constants of motion from a law of motion, or vice versa, but in the former
case we have to presume substantial laws, while in the latter case we
just presume the trivial fact that constants must exist. 
The latter presumption is sufficienty trivial to serve as a first 
principle, but we have a prize to pay: if the most fundamental constants 
are constants of motion that come along with some law of motion of some 
set of fundamental variables $\psi=(\psi_1,\psi_2,\dots,\psi_k)$, then 
a reference for a measurement of the components of $\psi$ is logically 
impossible. The fundamental level is a level of pure variables, while the 
``level'' of constants of motion is at least second order in $\psi$. 
As we have shown in Ref.~\cite{qed_paper,osc_paper}, the constants of 
motion are the even moments of $\psi$. Since COMs are required to obtain 
observables, then, within a \FiT, the fundamental level of reality 
can not be observed: The final theory must distinguish between existence 
and appearance.

In Ref.~\cite{qed_paper} we used an ``ontology of time'' as the basis 
for the same reasoning. But as we have shown in this essay, the 
suggested principle of variation (POV) can be derived from the idea 
of a \FiT, hence finally from the principle of sufficient reason. 
A genuine ontology is not required, as we derived the law, that all 
variables on some fundamental level have to vary, from distinct 
formal considerations.

Besides the distinction of variables from constants, we have to distinguish 
{\it quantity} from {\it structure}. It is trivially true that physical 
models contain variables that represent quantities. But structures 
have to be derived. A quantity can be represented by a number and a unit. 
A structure is given by an algebra. The complex plane for instance is 
represented by an algebra: The unit imaginary $i$ can not be represented 
by a number and {\it therefore} we (have to) introduce a symbol. 
Something similar holds for the number $\pi$, namely the fact that we can 
not write it down explicitely, suggests the use of a symbol.
But unlike $\pi$, which can by no means be completely written down,
we can represent the unit imaginary by a structure that fully 
characterizes the algebra of the complex numbers, namely that of 
$2\times 2$-matrices~\cite{osc_paper}. The introduction of 
a new symbol like $i$ is equivalent to the introduction of an algebraic 
structure. It is therefore reasonable to ask, if the unit imaginary is 
essential for quantum mechanics or not - and if it is, then why? 
The use of complex numbers in quantum mechanics can be questioned and 
hence it is a substantial choice. Since a \FiT may not presume what it 
is supposed to derive, it is not legitimate to simply presume a prominent 
role for complex numbers in a \FiT; the \FiT has to provide a reason.

This prelimenary survey does not claim to be complete. There are likely
more trivial principles that belong to the form of physics which we did not 
even mention. One of the principles that we did not discuss is implied in 
the possibility of a {\it quantitative} description: If there are observable 
quantitative properties of objects, then these properties must be additive: 
The length is of two rods is twice the length of a single rod. And there are
certainly more principles of this kind. We summarized these principles in the 
formula ``a world that allows for a physical description'', which is both,
precise enough for the purpose of this paper and vague enough to include
those principles that we did not discuss explicitely.

Let us finally emphasize that the describes principles imply a strong
selection: Whatever might be ``out there'', we can only measure those 
things that obey a law of motion which provides the necessary constants 
of motion as reference standards. Everything else is noise. Scientists 
should in general avoid to make claims about unmeasurable things. However, 
as we argued, with respect to a final theory, we can not avoid to make at 
least one claim about the wave-function, namely that its components are 
not directly measurable. This is not unreasonable, but a logical necessity.

\section{Objectivity and Classical Realism}
\label{sec_realism}

As explained in the previous sections, it appears that, given a final theory
would be possible or even the only logical consequence of the program of
some specific rationalism, then this seriously questions that objectivity 
and realism are compatible with each other; if the final theory would be found
and could be regarded as objectively and evidently (trivially) correct, then 
physical appearance and existence must be distinguished and classical realism 
must be regarded as disproven. Or, if we choose to believe that classical 
realism must be correct, then a final theory is impossible and there must be 
something {\it unreasonable}, i.e. some kind of significant but unreasonable
arbitariness in the world, something outside the PSR, something that a \FiT
could not derive or explain. Hence it appears that there is an intrinsic 
contradiction between realism and rationalism. The rationalist would expect 
that a \FiT is possible - but then a ``realistic'' worldview is logically
impossible.

We defined the notion of {\it objectivity} of experimental physics and 
we believe it is as clear as it is practically and theoretically relevant, 
though it refers to the notion of object, i.e. to something that is given 
only through {\it experience}. 
{\it Realism}, the complementary notion of {\it objectivity}, on 
the other hand, is more difficult to specify, also, because is appears in 
a number of variants. Often {\it realism} and {\it objectivity} are used in 
combination as {\it objective reality}, some sort of {\it common sense}
agreement about what is meant by ``the world''. Instead of making attempts to
give a theoretical definition of ``realism'', let us instead refer to a 
number of statements, that a strong realist would likely support:
\begin{itemize}
\item The world is in a fundamental way as we perceive it, i.e. composed of
 objects located in space and time. Objects are made of {\it matter}.
\item Every physical effect has an exclusively physical cause: The physical
world is {\it closed}. This includes:
\item The only possible way to influence the physical world is by physical
action (i.e. through actions of the body).
\item All valid information about the world originates from information 
that has been received physically, i.e. with finite velocity $v\le c$ via 
physical signals through our senses.
\end{itemize}
It is noteworthy, that while the principle of objectivity {\it defines} 
(and therefore limits) the scope of empirical sciences, especially that of
physics, realism defines (and therefore limits) reality, most often 
to the notions of space, time, matter and causality, or by 
suggesting limits of possible information transfer. The former leaves the
question open if the objectively real is a subset of the real, while
the latter reduces the world to the objective {\it by definition}. Strange
enough that these notions are rarely distinguished by physicists.
We do not intend to discuss this is much detail, but we would like to emphasize 
that the principle of objectivity defines a fundamental rule for quantitative 
empirical science that can not be questioned on scientific grounds, while 
realism tries to establish a specific ideology that is based to a large 
degree on unproveable metaphysical claims. One might summarize this as 
follows: it is logically impossible to question that physical 
laws are - within their limits of validity - {\it objectively true}, 
unless one questions that a kilogramm, a meter rod or a clock can be 
used as measurement standard. But classical realism reverses this logic 
and claims that {\it only} the ``objective'' aspects of the world ``really'' 
exist. We think that this type of restrictive realism is not only not
required, but it builds a barrier between natural and social sciences
that is not helpful. Besides that, the metaphysical assumptions of realism 
are not required to obtain objective knowledge of nature.

Especially the {\it closure} claim of classical realism can be considered 
problematic, especially in the context of the questions of free will and 
the nature of consciousness. 

\section{Some Formal Considerations}
\label{sec_form}

While we measure the weight of an object, by comparing it to reference weights,
we do not bother too much about the number system used. This might be deprecated
by pure mathematicians, but for most experimental physicist (and likely for 
most engineers), there are only two types of numbers required: integers and 
reals. The question, whether it is logically necessary to use the reals 
instead of the rational numbers, can be pragmatically answered insofar as 
there is no immediate reason to restrict the number system to the rationals, 
despite the fact that we can never obtain a non-rational measurement result 
in measurements of limited precision.
Though we can not measure $\pi$ with {\it infinite} precision, we are still 
be able to find ways to measure $\pi$ to {\it any required} precision by the
use a sufficiently large circles. Insofar the experimental determination of
$\pi$ via measurement is {\it in principle} not different from its
computation: No one can compute (or write down or store on a computer disc)
the number $\pi$ with {\it infinite} precision; but it can be done with any
required precision: there is no {\it logical} limit to the precision with 
which we can either compute {\it or} measure $\pi$.

Sometimes it is argued that the reals are - besides the complex numbers and
the quaternions - just the simplest field over which one can define {\it
 division rings} and that this fact suggests that the most general
mathematical objects (i.e. the quaternions) should be used on the fundamental
level of a final physical theory. Despite the fact, that quaternions might
play a prominent role in some areas of physics, we reject the general argument.
Both the complex numbers as well as the quaternions, can be represented 
by real valued structures, namely the complex numbers by real $2\times 2$ 
and the quaternions by real $4\times 4$ matrices.
Therefore they are not more fundamental than real matrices, but less, 
as they refer to specific matrix algebras. Instead of presuming a fundamental
role for a specific algebraic structure, a \FiT should derive and explain this
structure and prove its fundamentality. The intransparent mixture of quantity
and structure however may lead to confusion about the meaning and role of 
complex numbers in quantum mechanics and physics in general. This is indicated
by the discussion about the meaning of complex numbers in QM that is
ongoing for decades~\cite{Strocchi,Hestenes,BHW,GKS,Barbour,Rosales,MMG,Lev,CA,Mis,Kwong,CIA,Gibbons}.

We do not recognize the {\it progress} in mixing quantity and structure on the 
fundamental level. Rather, if complex numbers are a preferable, this should
be a result of the algebraic properties of the \FiT. 
This fact is reflected by the architecture of number systems
in computers, computers which are able to perform simulations of quantum
systems, i.e. integers and reals. A computer is usually understood
as purely based on binary digits. Though this is trivially true for 
contemporary {\it digital} computers, the existence of {\it analogue}
computers and the fact that the personal computer had originally two
processing units, the {\it central processing unit} (CPU) and the 
{\it floating point unit} (FPU), can be regarded as an argument against
the fundamentality of the {\it bit}: Even, if the memory of modern computers 
is by design based on {\it bits}, it is undeniable that we can not use 
it in a general scientifically meaningful way without introducing a 
{\it representation} of floating point numbers. This representation is 
based on bits for purely practical reasons. Though any given coding 
scheme of floating point numbers allows only for a limited 
precision, we are able to construct coding schemes for {\it any required 
precision}. 

\subsection{Structure Preservation}

{\it Any finite} physical system can be described by a (possibly infinite)
set of varying real fundamental quantities as a function of time $\psi(t)$. 
But all observable (measurable) properties are of even (at least
second) order in these fundamental variables. The variables themselves
can not be directly measured due to the lack of constant references.
References for measurements are constants of motion and they 
are at least of second order in $\psi$. Once we presume some (arbitrary) 
constant of motion ${\cal H}(\psi)$, then further constants of motion emerge 
in the description of this fundamental ``objects''~\cite{qed_paper,osc_paper}, 
namely the traces of powers of the autocorrelation matrix (matrix of second moments),
multiplied by the symplectic unit matrix. When the general constant of motion
${\cal H}(\psi)$ can be written as a (multivariate) Taylor series, then the
leading order terms automatically generate structure preservation. The presumed
constancy of ${\cal H}(\psi)$ generates a self-identity of ``objects'' 
via the identity of motional patterns. No other identity is 
required and possible. As we argued in Ref.~\cite{qed_paper,osc_paper}, the 
most general framework for structure preserving motion is {\it symplectic} motion
and it is a logical consequence of the constancy of ${\cal H}$~\footnote{
It can be shown that {\it unitary} motion is a special case of symplectic 
motion~\cite{Ralston}}. 
Furthermore it has been shown that the most simple, fundamental, and non-trivial 
algebra that describes the general properties of symplectic motion, is 
given by the real Dirac (Clifford-) algebra $Cl(3,1)$ which implies that 
the fundamental set of observables suggests the pattern of Minkowski 
space-time. There are $10$ sympletic generators within this Lie-algebra that
require an interpretation: a $3+1$ dimensional vector and a six-component 
bi-vector. The transformation properties of these components suggest
the interpretation in terms of energy, momentum, electric- and magnetic 
field~\cite{qed_paper,osc_paper} and the equations of motion of the
autocorrelation matrix allow for a derivation of the Lorentz force. 
However, the {\it primary and fundamental} variables are not space-time
coordinates any more. Instead the fundamental variables are the (abstract)
components of $\psi$, which can best be described as points in an abstract
phase space. These ideas thus reverse the role between ``real'' space and
phase space: The $3+1$-dimensional ``energy-momentum space'' is an effective 
space that emerges on the basis of a fundamental $4$-dimensional phase space 
described in proper time and the concept of space-time emerges through 
Fourier transformation. Effectively this Fourier transformation is similar 
to the characteristic function of probability theory, but its argument is 
not the probability density function (PDF) $\rho$, but the product of 
$\psi$ and $\sqrt{\rho}$.   
The $6$-dimensional phase space of a ``mass point'' emerges from the
equations of motion of the second moments and is symplectic only
in a low-energy and low-field approximation. 

Hence it seems possible to derive the concepts of mass and spin (the two
invarients of motion of a 4-spinor), of energy-momentum 4-vector, of the 
electromagnetic field and the transformation properties of the Lorentz group 
(rotations and boosts) on the basis of trivial principles. This is a first 
``proof of principle'' that a \FiT might indeed be possible - as a tautological 
theory.

\section{Summary}
\label{sec_summary}

In this paper we picked up the idea of a final theory of physics. First
we formulated the requirements of a final theory. These seem to be extreme
in view of the severe difficulties of modern physics to unify gravity and
quantum mechanics. But regardless of the current status of theoretical physics,
the idea of a final theory has its roots in the principle of sufficient reason 
(PSR) and hence essentially in rationalism. Specifically the idea that a
final theory would be inevitable and that it would carry the evidence for its
truth in itself seems to suggest that - following Goldberg's characterization 
of mathematics - a final theory of physics must be a tautology. This implies
that all statements of the theory are predetermined in its premises such that
it is possible to derive the content of physics logically from the premises, 
but also to elaborate to a higher degree of precision what a final theory is 
able to tell us about the world, and even more, what it is {\it not} able to 
tell. The absence of metaphysical presuppositions is a logical 
necessity of the finality of a \FiT - but it does not imply the 
absence of ``metaphysical'' conclusions. The final theory as we scetched it, 
might tell a lot about reality, but likely not what some of us expected to find. 
It will not tell us, why there is something instead of nothing. It will not 
even tell us, what type of substance matter ``really'' is. 
The tautological \FiT - and we claim that no other \FiT is possible - is, 
well, tautological.

Besides the PSR, which provides the rationale for a final theory, 
and is therefore obligatory, we discussed some premises for a \FiT that may
in summary be called the {\it form of physics}. We presented a first primitive 
analysis of the conjectured possibility of a \FiT. We identified three types 
of physical constants, namely dimensionless constants, dimensional conversion
factors and natural units. We explained why the only possible dimensional 
physical constant within a \FiT is the natural unit and that it can only 
emerge from the theory as a constant of motion.
We argued that this result logically implies that the fundamental level of 
a \FiT is not directly measurable - not because the fundamental level is less 
``real'', but because there is an unavoidable lack of reference.


\begin{thebibliography}{9}
  \section*{References}
\bibitem{Weinberg} Steven Weinberg; ``Dreams of a final theory'', Vintage
  Books (1993), ISBN 9780099223917.
\bibitem{ToE2005} G. t'Hooft, L. Susskind, E. Witten, M. Fukugita,
  L. Randall, L. Smolin, J. Stachel, C. Rovelli, G. Ellis, S. Weinberg and R. Penrose;
 Nature Vol. 433, 20th Jan. 2005, pp. 257-259.
\bibitem{Einstein20} A. Einstein in: The Living Age, Vol. 304 (3 Jan 1920), 41-43.
\bibitem{WikiReal} Wikipedia on ``Reality'' (Jan. 2017): https://en.wikipedia.org/wiki/Reality.
\bibitem{Tegmark} Max Tegmark: ``The Mathematical Universe'' (2007); arXiv:0704.0646.
\bibitem{Smilga} W. Smilga ``Reverse Engineering Approach to Quantum
    Electrodynamics''; J. of Mod. Phys. Vol.4 (2013), 561-571.
\bibitem{Reason} ``Principle of Sufficient Reason'', Stanford Encyclopedia of Philosophy; 
  https://plato.stanford.edu/entries/sufficient-reason/
\bibitem{PSR} Sir William Hamilton ``Lectures On Logic Vol. 1'', Boston: Gould (1859),
  https://archive.org/details/lecturesonmetaph00hamiuoft
\bibitem{WikiCave} Wikipedia on ``Allegory of the Cave'',
  https://en.wikipedia.org/wiki/Allegory\_of\_the\_Cave, dated August 14th, 2016.
  Original: Plato {\it Republic}, 514a.
\bibitem{Sokrates} Wikipedia on ``Sokrates'',
  https://en.wikipedia.org/wiki/Socrates, original: Plato {\it Apologeia},
  21d: ``Denn es mag wohl eben keiner von uns beiden etwas T\"uchtiges oder
Sonderliches wissen; allein dieser meint etwas zu wissen, obwohl er nicht
weiss, ich aber, wie ich eben nicht weiss, so meine ich es auch nicht.''.
  \bibitem{EinsteinBio} A. Einstein, {\it Autobiographical Notes}, Trans. by
    Paul Arthur Schilpp, Open Court Publ. 1979.
  \bibitem{Modinos} Antonis Modinos; ``From Aristotle to Schr\"odinger'',
    Springer (Cham, Heidelberg, New York, Dortrecht, London) 2014.
\bibitem{Boccaletti} Dino Boccaletti, ``Galileo and the Equations of Motion''
  Springer (Cham, Heidelberg, New York, Dortrecht, London) 2016.
\bibitem{Wiki} Wikipedia; https://en.wikipedia.org/wiki/Physical\_constant, 
(5th July 2016).
\bibitem{DiracLNH} P.A.M. Dirac ``The Large Numbers Hypothesis'' in: 
The Physicist's Conception of nature; Edt. Jagdish Mehra (1973), 
D. Reidel Publishing Company Dordrecht-Holland / Boston-U.S.A..
\bibitem{Hsu} Hsu, L. and Hsu, J. P. "The physical basis of natural units and truly
fundamental constants" (2012). Eur. Phys. J. Plus, 127: 11.
\bibitem{RalstonHBAR} John P. Ralston ``Quantum Theory without Planck's
  Constant'', arXiv:1203.5557, (2012).
\bibitem{Goldberg} Stanley Goldberg ``Understanding Relativity'', Birkh\"auser
  Basel Boston Stuttgart (1984). 
\bibitem{rdm_paper}  C. Baumgarten ``Use of real Dirac matrices in coupled linear optics''; 
Phys. Rev. ST Accel. Beams. 14, 114002 (2011).
\bibitem{qed_paper} C. Baumgarten; ``Relativity and (Quantum-) Electrodynamics 
from (Onto-) Logic of Time''; in Ruth E Kastner, Jasmina Jeknic-Dugic and George Jaroszkiewicz
  (Edt.) ``Quantum Structural Studies'', World Scientific (2017), ISBN: 978-1-78634-140-2. 
Preprint arXiv:1409.5338v5 (2014/2015).
\bibitem{osc_paper} C. Baumgarten; ``Old Game, New Rules: Rethinking the Form
  of Physics'', Symmetry 2016, 8(5), 30; doi:10.3390/sym8050030.


\bibitem{Strocchi} F. Strocchi, ``Complex Coordinates and Quantum Mechanics'',
Rev. Mod. Phys. 38, No. 1 (1966), pp. 36-40.
\bibitem{Hestenes} D. Hestenes ``Vectors, Spinors and Complex Numbers in Classical and Quantum Physics'',
Am. J. Phys. Vol. 39/9 (1971), 1013-1027.
\bibitem{BHW} W.E. Baylis, J. Huschlit and Jiansu Wei ``Why i?''
Am. J. Phys. Vol. 60 (1992), 788-795.
\bibitem{GKS} Philip Goyal, Kevin H. Knuth and John Skilling ``Origin of complex quantum amplitudes and Feynman's rules'',
Phys. Rev. A 81, 022109 (2010).
\bibitem{Barbour} Julian B. Barbour, ``Time and complex numbers in canonical quantum gravity'',
Phys. Rev. D 47, No. 12 (1993), pp. 5422-5429.
\bibitem{Rosales} J.L. Rosales ``Remarks on the issue of time and complex numbers in canonical quantum gravity'',
Phys. Rev. D 54 No. 6 (1996), pp. 4185-4188.
\bibitem{MMG} Matthew McKague, Michele Mosca and Nicolas Gisin ``Simulating Quantum Systems Using Real Hilbert Spaces'',
Phys. Rev. Lett. 102, 020505 (2009).
\bibitem{Lev} Felix Lev ``Why is Quantum Physics Based on Complex Numbers?'';  
Finite Fields and Their Applications, Vol. 12, 336-355 (2006), preprint: arXiv:hep-th/0309003.
\bibitem{CA} Charis Anastopoulos ``Quantum vs stochastic processes and the role of complex numbers'';
  Int.J.Theor.Phys.42:1229-1256,2003, preprint: arXiv:gr-qc/0208031.
\bibitem{Mis} J.A. Miszczak ``States and channels in quantum mechanics without complex numbers'',
arXiv:1603.04787.
\bibitem{Kwong} C.P. Kwong ``The mystery of square root of minus one in quantum mechanics, and its demystification'',
arXiv:0912.3996.
\bibitem{CIA} James M. Chappell, Azhar Iqbal, Derek Abbott ``Geometric Algebra: A natural representation of three-space'',
arXiv:1101.3619v4.
\bibitem{Gibbons} G. W. Gibbons ``The emergent nature of time and the complex numbers in quantum cosmology'',
arXiv:1111.0457v1.

\bibitem{Ralston} J.P. Ralston: ``{\cal PT} and {\cal CPT} quantum mechanics embedded in symplectic quantum mechanics'';
     J. Phys. A Math. Theor. Vol. 40 (2007), pp. 9883-9904.

\end{thebibliography}
\end{document}